\documentclass[11pt]{article}
\usepackage[utf8]{inputenc}
\usepackage[T1]{fontenc}
\usepackage{lmodern}
\usepackage[margin=1in]{geometry}
\usepackage{amsmath,amssymb}
\usepackage{graphicx}
\usepackage{booktabs}
\usepackage{hyperref}
\usepackage{xcolor}
\usepackage[numbers]{natbib}
\usepackage{enumitem}
\usepackage{placeins}
\usepackage{url}
\usepackage{microtype}
\hypersetup{colorlinks=true, linkcolor=blue, citecolor=blue, urlcolor=blue}

\title{One Index for Subsumption and Roll-up across Time, Geography, and Ontology}
\author{Madhulatha Mandarapu\thanks{madhulatha@samyama.ai} \and Sandeep Kunkunuru\thanks{sandeep@samyama.ai}}
\date{VaidhyaMegha Private Limited, India\\[2pt]\url{https://samyama.ai/}\\[8pt]June 2026}

\begin{document}
\maketitle

\begin{abstract}
Time-series, geospatial, and ontology systems each maintain a \emph{hierarchy} ---
\texttt{day\,$\sqsubseteq$\,month\,$\sqsubseteq$\,year}, \texttt{zip\,$\sqsubseteq$\,city},
\texttt{is-a}/\texttt{part-of} --- and each indexes it in a separate silo. We observe these are all
\emph{subsumption posets}, with one recurring workload: order testing (is $x$ under $y$?) and
hierarchical roll-up (aggregate a measure over everything under $y$). We present \textbf{OEH}, a single
declarable index that, by a cheap structural probe, encodes a hierarchy as a \emph{nested-set}
order-embedding (trees) or a \emph{chain decomposition} (low-width DAGs), and answers both subsumption
and \emph{index-resident} monoid roll-up from one structure. On five real hierarchies --- Gene Ontology,
NCBI Taxonomy (1.3M), GeoNames (330k), a 2.6M-node calendar, and git commit DAGs --- OEH on trees matches
a 2-hop index on query latency using $\sim$half the space and building 6--7$\times$ faster, and adds
roll-up that 2-hop cannot. Its roll-up matches TimescaleDB's continuous aggregates \emph{exactly} and in
the same latency regime, while also answering subsumption. On high-width DAGs the chain index is declined
and 2-hop dominates. Order-embedding is classical; our contribution is the unification and the
structure-selected index over subsumption \emph{and} index-resident roll-up.
\end{abstract}

\section{Introduction}
Three of the most active corners of data management each revolve around a \emph{hierarchy}, yet each
indexes it in isolation. Time-series engines roll up \texttt{minute$\to$hour$\to$day$\to$month} with
continuous aggregates; spatial systems nest \texttt{cell$\to$region$\to$country} with grids and
space-filling curves (S2, geohash); knowledge graphs and reasoners walk \texttt{is-a}/\texttt{part-of}
taxonomies. The structures and the queries are, underneath, the same: a partial order given by a
\emph{subsumption} (parent) relation, over which we ask (i) \emph{order tests} --- is $x$ subsumed in
$y$? --- and (ii) \emph{roll-up} --- aggregate a measure over everything subsumed by $y$.

We call such a structure a \emph{relationship hierarchy} and ask whether one index can serve all three
domains. The answer is a qualified yes. We present OEH (Order-Embedded Hierarchy), a single index that
chooses its encoding from the data's shape and answers subsumption and roll-up from one structure.

\paragraph{Contributions.}
\begin{itemize}
\item[C1] The relationship-hierarchy abstraction and its query algebra (\S\ref{sec:model}).
\item[C2] \textbf{OEH}: one declarable, structure-selected index answering subsumption \emph{and}
  \emph{index-resident} monoid roll-up --- the aggregate is answered \emph{from} the structure (a Fenwick
  range-sum over a nested-set, or a per-chain suffix-sum), not delegated to an engine join-group-aggregate.
  This is the precise distinction from the index-\emph{assisted} hierarchy line of SAP HANA
  (\S\ref{sec:related}).
\item[C3] An empirical study on five real hierarchies, including an exact cross-validation of roll-up
  against TimescaleDB and of subsumption against \texttt{git merge-base} (\S\ref{sec:eval}).
\item[C4] A regime map --- when nested-set, chain, or 2-hop each win --- with a principled
  $\sim\!8\sqrt{n}$ width cap.
\end{itemize}
We are deliberately scoped: OEH is a \emph{static} index (dynamic maintenance is future work), and
order-embedding/dominance as a subsumption primitive is classical prior art that we build on, not claim.

\section{Problem and model}\label{sec:model}
A \emph{relationship hierarchy} is a labeled partial order $(V,\sqsubseteq,\lambda)$ generated by a
covering (``parent'') relation; $x\sqsubseteq y$ reads ``$x$ is subsumed in $y$''. The query algebra has
two halves. \textbf{Order:} \textsf{subsumes}$(x,y)$, \textsf{ancestors}/\textsf{descendants}$(x)$,
\textsf{lca}. \textbf{Aggregation:} \textsf{rollup}$(\textit{measure},\ell)$ folds a monoid measure over
$\{y\}\cup\textsf{descendants}(y)$ at a target level $\ell$, with \emph{set semantics} on DAGs (each
descendant counted once). Time, geo, taxonomy, and genealogy are instances; the same two halves recur.

\section{The OEH index}\label{sec:method}
A cheap structural probe (the ``knob'') picks the encoding.

\paragraph{Trees $\to$ nested-set.} A DFS assigns each node an interval $[\textsf{in},\textsf{out}]$;
$x\sqsubseteq y \iff \textsf{in}(y)\le\textsf{in}(x)\wedge\textsf{out}(x)\le\textsf{out}(y)$ (2-D
containment). The subtree of $y$ is the contiguous \textsf{in}-order range $[\textsf{in}(y),\textsf{out}(y)]$,
so roll-up is a \textbf{Fenwick range-sum in $O(\log n)$}; two integers per node.

\paragraph{Low-width DAGs $\to$ chain decomposition.} A path partition gives each node a $(\textit{chain},
\textit{pos})$; for each node $v$ and chain $c$ we store $\textsf{reach}[v][c]$, the minimum reachable
position on $c$ (Jagadish's chain index~\cite{jagadish1990}); the chain count relates to the poset
width~\cite{dilworth1950}. Subsumption is $O(\textit{width})$. Crucially,
the descendants of $v$ on chain $c$ are the contiguous \emph{suffix} from $\textsf{reach}[v][c]$, so
set-semantics roll-up is $\sum_c \textsf{suffix-sum}_c$ --- exact and double-count-free.

\paragraph{Width cap.} Chain space is $O(n\cdot\textit{width})$; it only beats a 2-hop index while width
is small. OEH declines chain mode above $\sim\!8\sqrt{n}$ (so chain space stays $\sim\!O(n^{1.5})$) and
defers to 2-hop, which is the right substrate for high-width DAGs (\S\ref{sec:eval}).

\paragraph{Index-resident roll-up.} In both encodings the partial aggregate is answered \emph{from} the
index in $O(\log n)$ (trees) or $O(\textit{width})$ (DAGs), not by an engine aggregation over the group.
This is the capability that distinguishes OEH from prior hierarchy indexes (\S\ref{sec:related}).

\section{Evaluation}\label{sec:eval}
\paragraph{Setup.} Five real hierarchies: Gene Ontology~\cite{geneontology} (\textit{go-basic}, 38{,}263
nodes / DAG, 51\% multi-parent); NCBI Taxonomy~\cite{ncbitaxonomy} Metazoa subtree (1{,}323{,}391 / tree);
GeoNames~\cite{geonames} administrative hierarchy (329{,}993 / tree); a 5-year per-minute calendar
(2{,}675{,}155 / tree); and git commit DAGs (\texttt{postgres}, 102{,}560 / tree, width 38;
\texttt{git/git}, 84{,}891 / DAG, width 14\%). Baselines: a brute-force oracle; exact transitive closure;
GRAIL~\cite{yildirim2010}; PLL~\cite{akiba2013}; TimescaleDB hierarchical continuous aggregates~\cite{timescaledb};
\texttt{git merge-base} as a subsumption ground truth. Correctness is exact; timings are pure-Python and
machine-specific, used for apples-to-apples comparison.

\begin{figure}[t]\centering
\includegraphics[width=\linewidth]{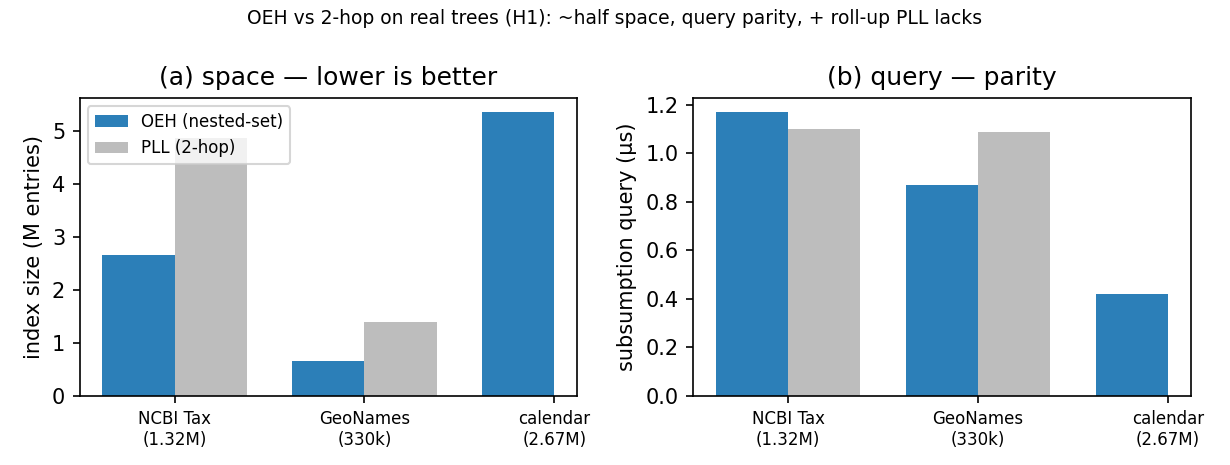}
\caption{H1: OEH nested-set vs.\ 2-hop (PLL) on real trees --- half the space, query parity.}
\label{fig:h1}
\end{figure}

\paragraph{H1: subsumption parity across three domains (Fig.~\ref{fig:h1}, Table~\ref{tab:h1}).}
One nested-set index serves ontology, geo, and time trees. Against PLL it uses $\sim$half the space and
builds 6--7$\times$ faster at query parity, and additionally answers roll-up PLL cannot.

\begin{table}[t]\centering\small
\caption{H1: OEH (nested-set) vs.\ PLL on real trees.}\label{tab:h1}
\begin{tabular}{lrrrrr}
\toprule
& $n$ & \multicolumn{2}{c}{space (M entries)} & \multicolumn{2}{c}{query ($\mu$s)}\\
\cmidrule(lr){3-4}\cmidrule(lr){5-6}
domain (dataset) & & OEH & PLL & OEH & PLL\\
\midrule
ontology (NCBI Tax) & 1{,}323{,}391 & 2.65 & 4.87 & 1.17 & 1.10\\
geo (GeoNames)      & 329{,}993    & 0.66 & 1.39 & 0.87 & 1.09\\
time (calendar)     & 2{,}675{,}155 & 5.35 & --- & 0.42 & ---\\
\bottomrule
\end{tabular}
\end{table}

\begin{figure}[t]\centering
\includegraphics[width=\linewidth]{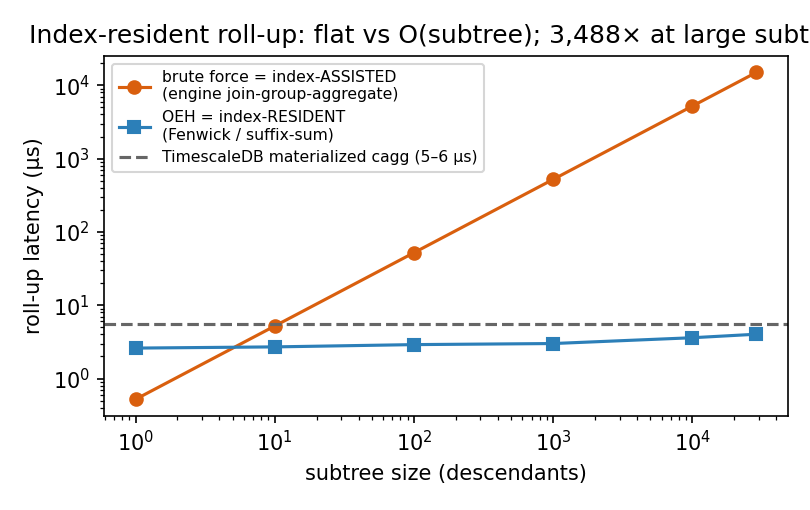}
\caption{Index-resident (OEH) vs.\ index-assisted (engine join-group-aggregate) roll-up.}
\label{fig:rollup}
\end{figure}

\paragraph{H2: index-resident roll-up (Fig.~\ref{fig:rollup}, Table~\ref{tab:ts}).}
Roll-up is exact and \emph{size-dependent}: OEH is $\sim$constant ($3$--$4\,\mu$s, $O(\log n)$) regardless
of subtree size, whereas an engine-style aggregation is $O(\textit{subtree})$. On large subtrees (avg
28{,}851 descendants) OEH is $3{,}488\times$ faster; it loses only on trivially small subtrees. The
brute-force baseline is precisely the index-\emph{assisted} join-group-aggregate of the HANA line
(\S\ref{sec:related}); OEH beating it by orders of magnitude is the direct case for index-residence.
Against TimescaleDB on the same calendar, OEH's roll-up matches \emph{exactly} (day $704{,}800$, month
$21{,}168{,}000$) and in the same single-digit-$\mu$s regime as a \emph{materialized} continuous
aggregate, while also answering subsumption, which a continuous aggregate cannot.

\begin{table}[t]\centering\small
\caption{Time-axis roll-up: OEH vs.\ TimescaleDB ($\mu$s). Sums match exactly.}\label{tab:ts}
\begin{tabular}{lrrr}
\toprule
level & OEH (index-resident) & TS cagg (materialized) & TS raw\\
\midrule
day   & 2.63 & 5.09 & 92.17\\
month & 3.13 & 6.04 & 1516.58\\
\bottomrule
\end{tabular}
\end{table}

\begin{figure}[t]\centering
\includegraphics[width=0.92\linewidth]{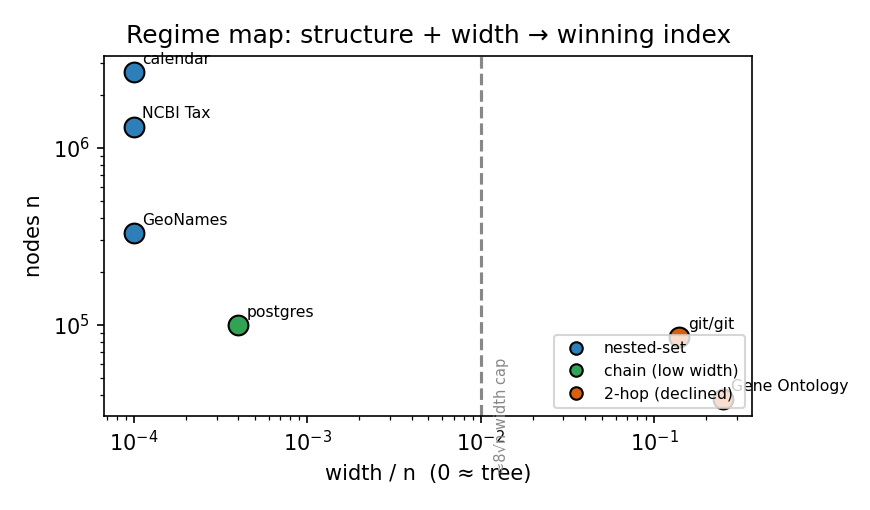}
\caption{Regime map: structure and width determine the winning index.}
\label{fig:regime}
\end{figure}

\paragraph{H3: regime map (Fig.~\ref{fig:regime}).} The knob picks nested-set for trees, chain for
low-width DAGs, and defers to 2-hop otherwise. On Gene Ontology (width $\approx$ its 22{,}807 leaves) and
on \texttt{git/git} (width 14\% of $n$) chain mode auto-declines and PLL owns the regime. On
\texttt{postgres} (a rebase history, width 38) forced chain is compact and correct; on \texttt{git/git}
forced chain is validated correct against \texttt{git merge-base} itself but not space-efficient. A
finding worth stating: genuinely low-width \emph{multi-parent} DAGs are rare --- real merge histories are
high-width, real low-width histories are trees.

\FloatBarrier
\section{Related work}\label{sec:related}
\paragraph{Order-embedding as subsumption.} Encoding a hierarchy so that subsumption becomes
coordinate-wise dominance is classical: tree ancestry as a pre/post plane indexed by an
R-tree~\cite{grust2002}, bit-vector lattice encodings~\cite{aitkaci1989}, and dominance-drawing
reachability~\cite{feline2014,ortali2021} where the number of coordinates equals the Dushnik--Miller
order dimension~\cite{dushnik1941}. These are \emph{static, reachability-only}; we build on them and do
not claim the primitive as novel.

\paragraph{Reachability labeling.} 2-hop~\cite{cohen2003}, tree-cover~\cite{agrawal1989},
GRAIL~\cite{yildirim2010}, PLL~\cite{akiba2013}, and the incremental, append-only,
width-parameterized index of Bulteau et al.~\cite{bulteau2025} answer reachability but support \emph{no}
aggregation. Bulteau et al.\ are dynamic by design (Merkle graphs); OEH is static and instead adds roll-up.

\paragraph{Hierarchy indexes with aggregation.} The TUM/SAP line is the closest: DeltaNI~\cite{finis2013}
(versioned nested intervals; ``aggregate queries out of scope''), Order Indexes~\cite{finis2015}, SAP
HANA hierarchies~\cite{brunel2015}, and the aggregation-focused Index-Assisted Hierarchical
Computations~\cite{brunel2016}. These \emph{do} roll up user measures, but \emph{index-assisted}: the
index answers \texttt{IS\_DESCENDANT}/order/level in $O(\log n)$ and a relational structural-grouping
(join-group-aggregate) operator computes the sum per query --- no measure or partial sum is stored in the
index. OEH's roll-up is \emph{index-resident}: the partial aggregate is answered from the structure, from
the same index that answers subsumption. Empirically (Fig.~\ref{fig:rollup}) their approach is our
brute-force baseline.

\paragraph{OLAP and dynamic theory.} Dimension-hierarchy models~\cite{pedersen2001}, cube
materialization~\cite{harinarayan1996}, and Graph Cube~\cite{zhao2011} roll up but do not unify roll-up
with a subsumption index over the value poset; the cuboid lattice is not the value-subsumption poset.
Dynamic-treewidth maintenance~\cite{korhonen2023} is relevant only to a future dynamic variant.

\section{Limitations and honest findings}\label{sec:limits}
\begin{itemize}
\item \textbf{Static only.} OEH supports no online insert/delete; on the dynamic axis~\cite{bulteau2025,
  finis2013} are ahead. We do not claim a dynamism advantage.
\item \textbf{Chain mode rarely wins on real DAGs.} Real multi-parent DAGs (ontologies, merge histories)
  are high-width, so chain declines and 2-hop wins; low-width multi-parent DAGs are uncommon.
\item \textbf{Roll-up loses on tiny subtrees} (the $O(\log n)$ constant exceeds $O(\textit{small})$).
\item The TimescaleDB comparison is in-process Python vs.\ SQL --- same regime, not a system benchmark.
\item Order-embedding, nested-set, and dominance are prior art (\S\ref{sec:related}).
\item GRAIL/PLL are re-implementations (validated exact vs.\ the oracle); GeoNames is a near-tree DAG
  (0.9\% multi-parent kept to one canonical parent); the chain decomposition is greedy and a near-minimum
  cover would be smaller.
\end{itemize}

\section{Conclusion}
Relationship hierarchies unify the hierarchies that time-series, spatial, and ontology systems index
separately. OEH is one structure-selected index that answers subsumption and index-resident roll-up
across all three domains, validated on real data and against a production time-series engine. The honest
boundary --- high-width DAGs defer to 2-hop --- and a future dynamic variant are the natural next steps.

\bibliographystyle{plainnat}
\bibliography{refs}

\end{document}